\documentclass[conference]{IEEEtran}
\IEEEoverridecommandlockouts
\ifCLASSINFOpdf
\else
\fi
\usepackage{lipsum}
\usepackage{textcomp}
\usepackage{graphicx}
\usepackage{float}
\usepackage{cite}
\usepackage{booktabs}
\usepackage{epstopdf}
\usepackage{multirow}
\usepackage{boldline}
\usepackage{amsmath}
\usepackage{relsize}
\usepackage{array}
\usepackage{colortbl}
\usepackage{algorithm}
\usepackage{algpseudocode}
\usepackage{placeins}
\usepackage{tabulary}
\usepackage{makecell}
\usepackage{enumitem}
\usepackage{mathtools}
\usepackage{soul}
\usepackage[normalem]{ulem}
\newcommand\soo{\bgroup\markoverwith{\textcolor{red}{\rule[0.5ex]{2pt}{0.4pt}}}\ULon}

\makeatletter
\newlength \figwidth
\if@twocolumn
\else
\fi
\makeatother
\makeatletter
\def\ps@IEEEtitlepagestyle{%
  \def\@oddfoot{\mycopyrightnotice}%
  \def\@evenfoot{}%
}
\def\mycopyrightnotice{%
  {\footnotesize \textit{This paper is accepted at the IEEE International Conference on Communications (ICC)} $\mathit{2018}$\hfill}% <--- Change here
  \gdef\mycopyrightnotice{}% just in case
}

\begin{document}
\IEEEoverridecommandlockouts
%\title{On the Stationarity of Realistic Wireless Body-to-Body Channels}
%\title{On the Existence of Stationarity in Realistic Wireless Body-to-Body Channels}
\title{Wide-Sense-Stationarity of Everyday Wireless Channels for Body-to-Body Networks}
%\title{Wide-Sense-Stationarity of Realistic Wireless Body-to-Body Channels}
%\title{On the Stationarity of Wireless Body-to-Body Channels: an Experimental analysis}
%\title{Stationarity Analysis of Realistic Wireless Body-to-Body Channels}

\author{\IEEEauthorblockN{Samiya~M.~Shimly}
\IEEEauthorblockA{The Australian National University$^\mathsection$\\
CSIRO Data61\\
Email: Samiya.Shimly@data61.csiro.au}\vspace{-20pt}
\and
\IEEEauthorblockN{David~B.~Smith}
\IEEEauthorblockA{CSIRO Data61\\
The Australian National University\\
David.Smith@data61.csiro.au}\vspace{-20pt}
\and
\IEEEauthorblockN{Samaneh~Movassaghi}
\IEEEauthorblockA{The Australian National University$^\mathsection$\\
CSIRO Data61\\
Samaneh.Movassaghi@data61.csiro.au}\vspace{-20pt}
\thanks{$^\mathsection$This research is supported by an Australian Government Research Training Program (RTP) scholarship.}}

%Telephone: (800) 555--1212, Fax: (888) 555--1212}

\maketitle
\begin{abstract}
The existence of wide-sense-stationarity (WSS) in narrowband wireless body-to-body networks is investigated for ``everyday" scenarios using many hours of contiguous experimental data. We employ different parametric and non-parametric hypothesis tests for evaluating mean and variance stationarity, along with distribution consistency, of several body-to-body channels found from different on-body sensor locations. We also estimate the variation of power spectrum to evaluate the time independence of the auto-covariance function. Our results show that, with 95\% confidence, the assumption of WSS is met for at most 90\% of the cases with window lengths of 5 seconds for the channels between the hubs of different BANs. Additionally, in the best-case scenario, the hub-to-hub channel remains reasonably stationary (with more than 80\% probability of satisfying the null hypothesis) for longer window lengths of more than 10 seconds. The short time power spectral variation for body-to-body channels is also shown to be negligible. Moreover, we show that body-to-body channels can be considered wide-sense-stationary over significantly longer periods than on-body channels.
%Moreover, CMR provides up to 9 dB performance improvement over SPR, with 90\% packet delivery ratio.
\end{abstract}
% \fi
\IEEEpeerreviewmaketitle
\section{Introduction}
Wireless body-to-body networks (BBNs) can enable coexistence of wireless body area networks (BANs) by exploiting body-to-body (B$2$B) communications using wearable on-body hub/sensor devices. While BANs are specifically designed to collect data from various sensors placed on/inside or around the human body, BBNs send data through closely located BANs to reach the intended destination/server in case of unavailable or out-of-range network infrastructure (in emergency indoor/outdoor situations) \cite{shimly2017cross}. BBNs are envisioned to be self-organizing, smart and mobile networks that can create their own centralized/decentralized network connection without any external coordination. This requires systematic prediction and modeling of the channel behavior. Statistical characterization of a channel requires time segments that possess wide-sense-stationarity (WSS) or second-order stationarity where the first and second moment (i.e., mean, variance and auto-covariance) of the channel are independent of time \cite{chaganti2014narrowband}. Also in \cite{bello1963characterization}, Bello suggested that, his proposed wide-sense-stationary uncorrelated scattering (WSSUS) assumption can only be held for limited intervals of time and frequency as the real-world radio channels often demonstrate `quasi-stationary' behavior. Therefore, it is important to estimate the channel parameters to identify the wide-sense stationary regions to see if these model parameters can be applied over a suitable time-frame.

To test  the WSS of wireless channels, a parametric approach is proposed in \cite{kay2008new} to detect non-stationarity based on the time-variant autoregressive (TVAR) model. A parametric unit-root test is proposed in \cite{reinsel2003elements} to parameterize a predetermined structure. Willink tested the WSS of multiple-input multiple-output (MIMO) wireless channels in \cite{willink2008wide} by investigating the first and second moment with parametric one-way ANOVA and non-parametric time-dependent evolutionary spectrum analysis, respectively. Other non-parametric approaches to identify the stationarity intervals include run-test described in\cite{bultitude2002estimating}, comparison of the delay power spectral density (PSD) estimated at different time instances \cite{umansky2009stationarity} and evaluation of the variation of time-localized PSD estimate \cite{basu2009nonparametric}.

However, BAN/BBN channels are practically different to the typical wireless/radio channels because of the slowly-varying human-body dynamics and shadowing caused by postural body movements \cite{smith2015channel}. Hence, in \cite{chaganti2014narrowband} the authors used different parametric and non-parametric approaches for testing WSS of on-body channels and showed that on-body channels have non-stationary characteristics. \textit{The novelty here is the investigation of whether the WSS assumption can be applied for body-to-body (B$2$B) channels and to find the typical duration for WSS regions of B$2$B channels.} We use parametric one-way ANOVA for investigating mean stationarity and non-parametric Brown--Forsythe (B--F) test and Kolmogorov--Smirnov (K--S) test to investigate variance stationarity and distribution consistency of the channels, respectively. We also use variation in the PSD estimate for testing the time independence of the auto-covariance of the channels. Our findings based on the application of the aforementioned tests on the experimental setup are as follows:
\begin{itemize}
\item For body-to-body channels, the hub-to-hub (Left-Hip to Left-Hip) links show better probability of satisfying the wide-sense-stationarity (WSS) assumption than the hub-to-sensor (i.e., Left-Hip to Right-upper-Arm, Left-Hip to Left-Wrist) links.
\item According to the tests, there is approximately up to $90\%$ probability (over the total period) that the hub-to-hub links will satisfy the null hypothesis for window lengths of $5$ s with $95\%$ confidence level (also up to $85\%$ for window lengths of $10$ s with $99\%$ confidence level).
\item In the best-case scenario, the hub-to-hub channel can satisfy the null hypothesis assumption with a window length of $50$ s for more than $85\%$ time over the whole period (with $95\%$ confidence level).
\item Negligible variation in power spectral density is found for different window lengths (e.g., $5$ s, $10$ s) amongst many different B$2$B channels.
\item Body-to-body links are more stationary with respect to on-body links, as on-body links show non-stationary behavior with $50\%$ chance of rejecting the null hypothesis over the whole period for an estimated minimum required window length of $3$ s.
\item Even in the best-case scenario, on-body links show lower probability of being stationary (tending to non-stationary behavior) than B$2$B links. \emph{Hence, from this analysis, in conjunction with on-body results in \cite{chaganti2014narrowband}, B$2$B communications shows significantly more stationarity than on-body communications.}
\item For B$2$B channels, the probability of satisfying WSS can depend on the sensor locations, as hub-to-hub and hub-to-sensor links show varying probability of satisfying the null hypothesis.
\item The probability of being stationary for all B$2$B channels decreases with increasing the window length.
\end{itemize}

\section{Experimental Scenario}
\begin{figure}[!t]
\centering{\includegraphics[width=60mm]{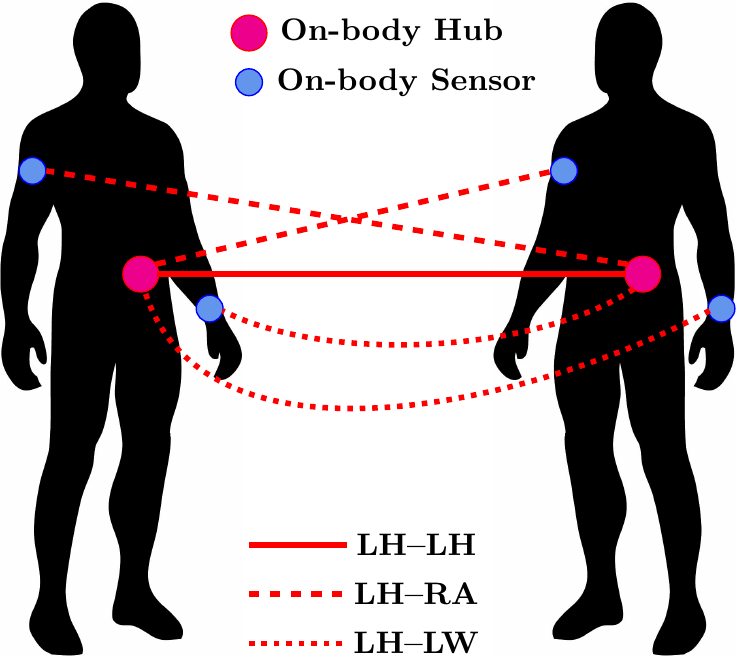}}
\caption{Different Body-to-Body links between two BANs wearing on-body hub at the left hip (LH) and two on-body sensors at the right upper arm (RA) and left wrist (LW), respectively.}
\label{b2b}
\end{figure}

We use an open-access dataset which consists of contiguous extensive intra-BAN (on-body) and inter-BAN (body-to-body) channel gain data of around $45$ minutes, captured from $10$ closely located mobile subjects (adult male and female)\footnote{Hence many hours of contiguous link data} with $50$ ms sampling time. The experimented subjects were walking together to a crowded hotel bar, remaining there for a while and then walking back to the office. Each subject wore $1$ transmitter (Tx hub) on the left-hip and $2$ receivers (sensors/ relays) on the left-wrist and  right-upper-arm, respectively (Fig. \ref{b2b}). The radios were transmitting at $0$~dBm power with $-100$ dBm receive sensitivity. A description of these wearable radios can be found in \cite{hanlen2010open} and the ``open-access" dataset can be downloaded from \cite{smith2012body}. Each Tx transmits in a round-robin fashion, at $2.36$ GHz, with $5$ ms separation between transmission and also acts as Rx capturing RSSI (Receive Signal Strength Indicator) values.

We investigate the WSS of three different body-to-body (B$2$B) links i.e., left hip to left hip (LH--LH), left hip to right upper arm (LH--RA) and left hip to left wrist (LH--LW) and average the results from $10$ BANs. An illustration of the B$2$B links found from different on-body sensor locations is shown in Fig. \ref{b2b} with two BANs. We also investigated single B$2$B links with good and bad conditions from $10$ BANs (shown in Table \ref{table_bw}) to investigate best-case and worst-case stationarity of different B$2$B links.

\begin{table}[!t]
\centering
\caption{Best and Worst Case Links over $10$ subjects}\vspace{-0.5em}
\label{table_bw}
\renewcommand{\arraystretch}{1.2}
\begin{tabular}{|c|c|c|c|}\Xhline{0.8pt}
 & \textbf{LH--LH} & \textbf{LH--RA} & \textbf{LH--LW}\\[1ex]\Xhline{1pt}
Best-case & BAN$7$ -- BAN$5$ & BAN$5$ -- BAN$7$ & BAN$7$ -- BAN$8$\\\hline
Worst-case & BAN$2$ -- BAN$9$ & BAN$2$ -- BAN$1$ & BAN$2$ -- BAN$10$\\\Xhline{0.8pt}
\end{tabular}
\end{table}
\section{Tests of Significance for WSS}
We use the ``frequentist" approach along with null hypothesis significance testing (NHST) \cite{nickerson2000null} to investigate the wide-sense-stationarity (WSS) of the B$2$B channels, with different test statistics (i.e., difference between mean, variance and distribution properties). We also examine the variation in power spectrum which gives the auto-covariance characteristic of the channels. Wide-sense-stationarity requires that the first and second moments (i.e., mean, variance, auto-covariance) of a time varying stochastic process $X(t)$ do not vary with respect to time $t$. In this paper, WSS is tested over a wide range of window lengths ($L$) from $100$ ms to $100$ s, such as, $L = [100,200,300,...,100000]$ ms. Here, we follow the process from \cite{chaganti2014narrowband}, where the whole channel is divided into $m$ consecutive non-overlapping intervals of length $\ell$ (where $\ell = L/2$) to perform $(m-1)$ independent pairwise comparisons across two consecutive intervals. Hence, for each window length $L$, i.e., ($L = 2\ell$), there will be $(m-1)$ pairwise independent null hypothesis tests. We estimate the average probability of stationarity for a window length of $L$ over $(m-1)$ tests with NHST for test statistic $T_L$ at a significance level of $\alpha$, where we consider:
\begin{equation*}
\begin{multlined}
\hspace{1em}H_0 \colon \textit{ $L$ retains WSS (null hypothesis)}\\
H_1 \colon \textit{ $L$ does not retain WSS (alternative hypothesis)}
\end{multlined}
\end{equation*}
Then,
\begin{equation}
p_L = P\big\{T_L \geq T_{L_{obs}} \bigl\vert H_0\big\}
\end{equation}
where $p_L$ is the probability of observing a more extreme test statistic ($T_L$) than the one observed ($T_{L_{obs}}$), given that the null hypothesis is true (observing a significant difference due to random sampling error while there was none or negligible difference).
\begin{equation*}
\textit{if } p_L \geq \alpha, \quad H_0 \textit{ is not rejected}
\end{equation*}
\begin{equation*}
\begin{multlined}
\textit{if } p_L < \alpha, \quad H_0 \textit{ is rejected in favor of } H_1
\end{multlined}
\end{equation*}
where $\alpha$ is the significance level/threshold for measuring the significance of the test outcome (based on $p_L$), which can be interpreted as the probability of incorrectly rejecting a true null hypothesis. We examine a range of statistical significance with $\alpha$ $\epsilon$ $\{0.01, 0.05, 0.1\}$, which corresponds to a confidence level ($c\ell$) of $c\ell$ $\epsilon$ $\{0.99, 0.95, 0.90\}$, as $c\ell = (1 - \alpha)$. For example, $\alpha = 0.05$ implies that, while there is $5\%$ probability of incorrectly rejecting the null hypothesis, there is $95\%$ probability that the confidence interval contains the null hypothesis value (e.g., $0$ for difference, $1$ for ratio) \cite{tan2010correct}.

\vspace{0.5em}The average probability of stationarity ($\gamma_L$) for a window length $L$ over the entire channel can be calculated as follows:
\begin{equation}
\gamma_L = \frac{\sum_{i=1}^{m-1}\big\{{p_L}^i \geq \alpha \big\}}{m-1}
\label{Prob_St}
\end{equation}
which also implies the average probability of satisfying the null hypothesis for window length $L$ over the whole period (from $m-1$ pairwise comparisons). When calculating the average ${p_L}^i$ for $i^{th}$ pairwise comparison ($i^{th}$ window) over multiple similar links from different subjects, we choose the median (typical) value (${\tilde{p_L}}^i$) to obtain a more robust estimation, as the median is not effected by outliers.

A brief description of the statistical hypothesis tests conducted here along with the grounds for choosing those tests are given in the  following subsections.

\subsection{ANOVA Test}
The ANOVA (Analysis of Variance) test is used for analyzing the variation (as the name implies) or difference between the means of two or more sets of observations. We use the parametric one-way ANOVA test statistic ($T_{L_{anova}}$) which is the ratio of the mean square variance between the intervals to the mean square variance within each interval \cite{scheffe1999analysis}.
\begin{equation}
T_{L_{anova}} = \frac{\bar{S}_{between}}{\bar{S}_{within}}
\end{equation}
where
\begin{equation}
\bar{S}_{between} = \frac{\sum_{i=1}^{m_t}n_i(\bar{X}_i - \bar{X})^2}{m_t - 1}
\end{equation}
and
\begin{equation}
\bar{S}_{within} = \frac{\sum_{i=1}^{m_t}\sum_{j=1}^{n_i}(\bar{X}_{ij} - \bar{X}_i)^2}{N - m_t}
\end{equation}
where $m_t$ is the number of intervals over which the hypothesis is being tested (here, $m_t = 2$), $X_{ij}$ is the $jth$ element of the $ith$ interval, $n_i$ is the number of observations in $ith$ interval and $N$ is the total number of observations across $m_t$ intervals. $\bar{X}$ is the mean over $m_t$ intervals $\big(\bar{X} = \frac{1}{N}\sum_{i=1}^{N}X_i\big)$.
This test relies on the assumption of the normality and homogeneity of the variances of the underlying distribution. In general, the B$2$B channels are not normally distributed (they typically possess a skewed distribution). Fortunately, ANOVA is fairly robust to moderate deviations from normality \cite{glass1972consequences,lix1996consequences}, specially with a large number of observations. Additionally, it is not very sensitive against the homoscedasticity (homogeneity of the variances) assumption with balanced data (when the sets/intervals are the same size and have similar distribution) \cite{statguide1997your}. Alternatively, a nonparametric version of the ANOVA (Kruskal-Wallis (K--W) test \cite{daniel1990kruskal}) can be used, which does not depend on the normality assumption. By comparing the results of the K--W test and ANOVA test, negligible difference was observed. Hence, the classical one-way ANOVA analysis results are provided here.

\subsection{Brown--Forsythe Test}
To investigate the homogeneity of the variances over the window lengths, hence further testing the homoscedasticity assumption made for ANOVA, we use the non-parametric Brown--Forsythe (B--F) test \cite{brown1974robust}, which calculates the $F$ statistic resulting from an one-way ANOVA on the absolute deviations from the median.
\begin{equation}
T_{L_{BF}} = \frac{\frac{\sum_{i=1}^{m_t}n_i(\bar{d}_i - \bar{d})^2}{m_t - 1}}{\frac{\sum_{i=1}^{m_t}\sum_{j=1}^{n_i}(\bar{d}_{ij} - \bar{d}_i)^2}{N - m_t}}
\end{equation}
where $d = \bigl\lvert X_{ij} - \tilde{X_i}\bigl\rvert$ and $\tilde{X_i}$ is the median of the $i^{th}$ interval.
This is a modified version of the Levene's test \cite{levene1960robust} (estimation of the deviation from the mean) which does not rely on the normality assumption, and therefore provides good robustness against many types of non-normal data while retaining good statistical power \cite{schultz1985levene,felix2015pervasive}. Also, non-parametric tests are more useful when investigating physical phenomena , e.g., radio propagation, as unlike parametric tests they make no assumptions regarding the probability distributions of the sampled process \cite{willink2008wide}.

\subsection{Kolmogorov--Smirnov Test}
We use the nonparametric two-sided Kolmogorov-Smirnov (K-S) two-sample test \cite{gibbons2011nonparametric} to examine whether the samples of two consecutive intervals come from the same distribution. This test is sensitive to any difference in median, dispersion and skewness between two distributions, as it estimates the maximum absolute difference between the two empirical distributions as follows,
\begin{equation}
T_{L_{KS}} = sup_{(x)}\bigl\lvert F_y(x) - F_z(x)\bigr\rvert, \qquad x = x_1,...,x_{n+l}
\end{equation}
where $T_{L_{KS}}$ is the K-S test statistic and $y = X(t_1),...,X(t_n)$ and $z = X(t_{1+l}),...,X(t_{n+l})$ are two consecutive intervals of $X(t)$, where $t_n$ is the element at time instance $t$.

\section{Test Results}
We estimate the average probability of stationarity over the whole period for different window lengths $L$, which implies the percentage of the pairwise comparisons over the total period for which a window length $L$ is satisfying the null hypothesis (WSS assumption). If the percentage is higher, then we consider the channel can possess WSS. The test outcome of $L$ is averaged over multiple links (e.g., $90$ links) with similar source and destination nodes chosen from different subjects (e.g., $10$ BANs) to get an approximation of the WSS property of the B$2$B links with varying distributions. We also examine several single B$2$B links with best-case and worst-case scenarios (listed in Table \ref{table_bw}) to get  an assumption of random B$2$B channel stationarity in different conditions. We have applied a varying range of interval length ($\ell = [50,100,150,...,50000]$ ms) with a sampling frequency of $20$ Hz, hence providing a good variation of the number of samples/observations ($\ell_n = [1,2,3,...,1000]$) within each interval, which contains the minimum number of samples (e.g., $30$, $50$) required for an interval according to \cite{wilks1997resampling,chaganti2014narrowband} to minimize the probability of Type-I and Type-II errors. Thus, to precisely investigate the WSS property of the channels, we consider window lengths of greater than or equal to $3$ seconds which contains intervals ($\geq$ $1.5$ s) with greater than or equal to $30$ samples.

\subsection{Mean Stationarity}
\begin{figure}[!t]
\centering{\includegraphics[width=\figwidth]{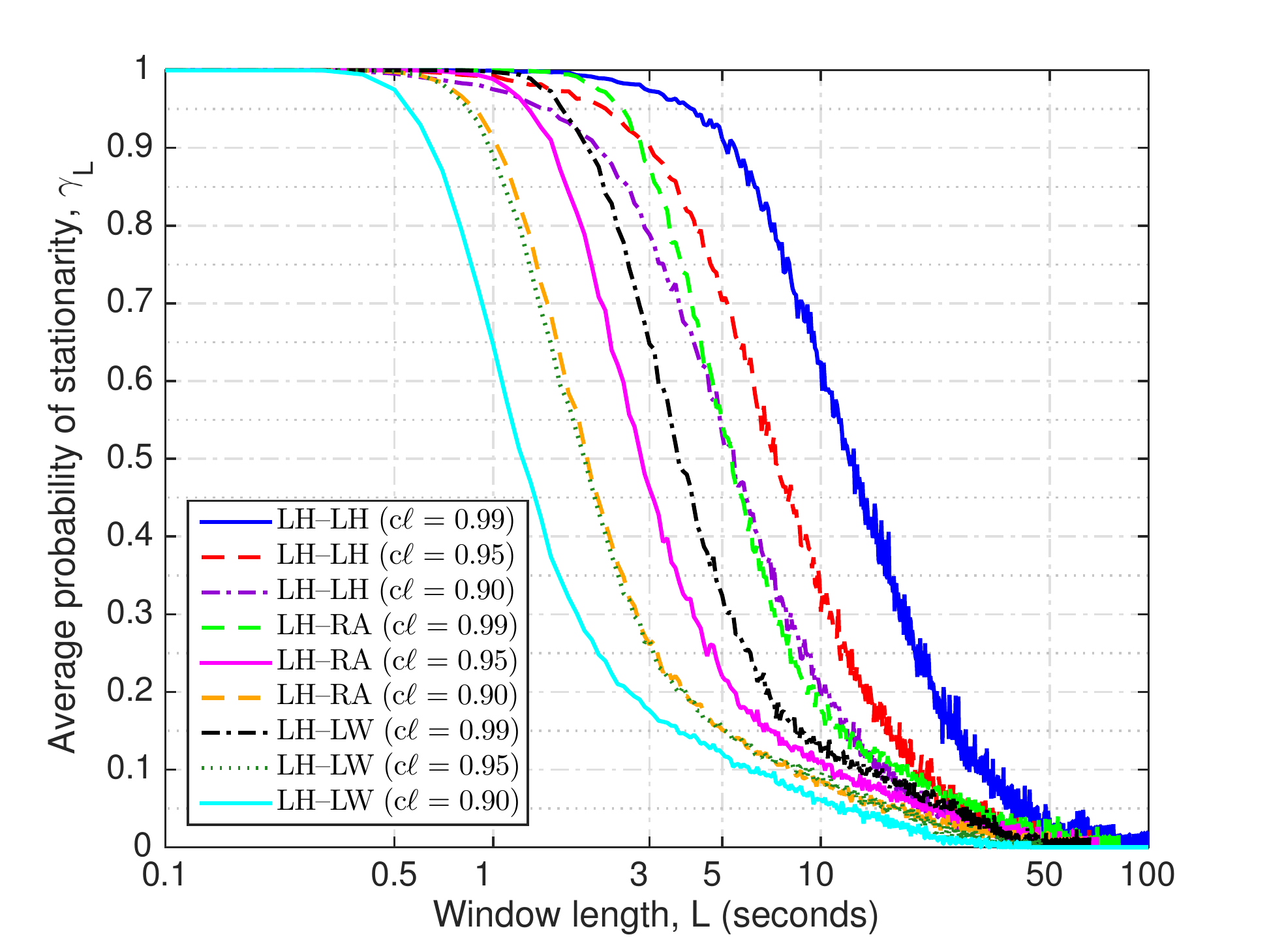}}
\caption{ANOVA hypothesis test for average probability of stationarity across different body-to-body links, i.e., L. hip to L. hip (LH-LH), L. Hip to R. upper Arm (LH-RA), L. Hip to L. Wrist (LH-LW) over $10$ subjects.}
\label{anova_all}
\end{figure}

The average probability of stationarity for the ANOVA hypothesis test over $10$ BANs with different body-to-body links is shown in Fig. \ref{anova_all}, where the hub-to-hub links have better stationarity than hub-to-sensor links between different BANs. For example, with a $99\%$ confidence level, the LH--LH link has more than $90\%$ and $60\%$ chance of satisfying the null hypothesis over the whole period with window lengths of $5$ s and $10$ s, respectively. However, with $95\%$ confidence level ($5\%$ chance of error), LH--LH links show $70\%$ probability of being stationary for $5$ s window length which goes down to $30\%$ for window length of $10$ s. Among the hub-to-sensor links, the LH--RA link shows better stationarity than the LH--LW link.

\subsection{Variance Stationarity}
\begin{figure}[!t]
\centering{\includegraphics[width=\figwidth]{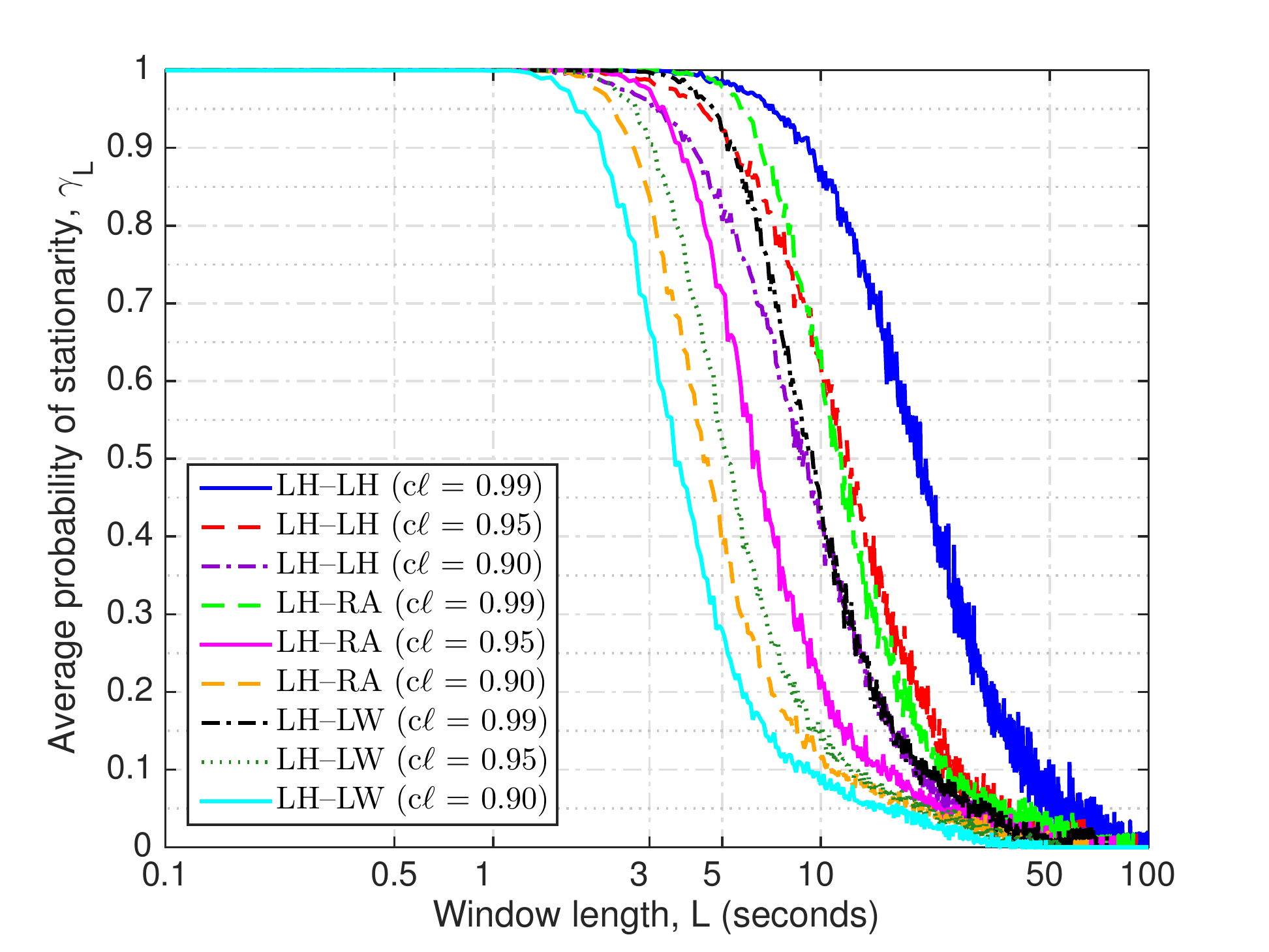}}
\caption{B--F hypothesis test for average probability of stationarity across different body-to-body links, i.e., L. hip to L. hip (LH-LH), L. Hip to R. upper Arm (LH-RA), L. Hip to L. Wrist (LH-LW) over $10$ subjects.}
\label{BF_all}
\end{figure}

\begin{figure}[!t]
\centering{\includegraphics[width=\figwidth]{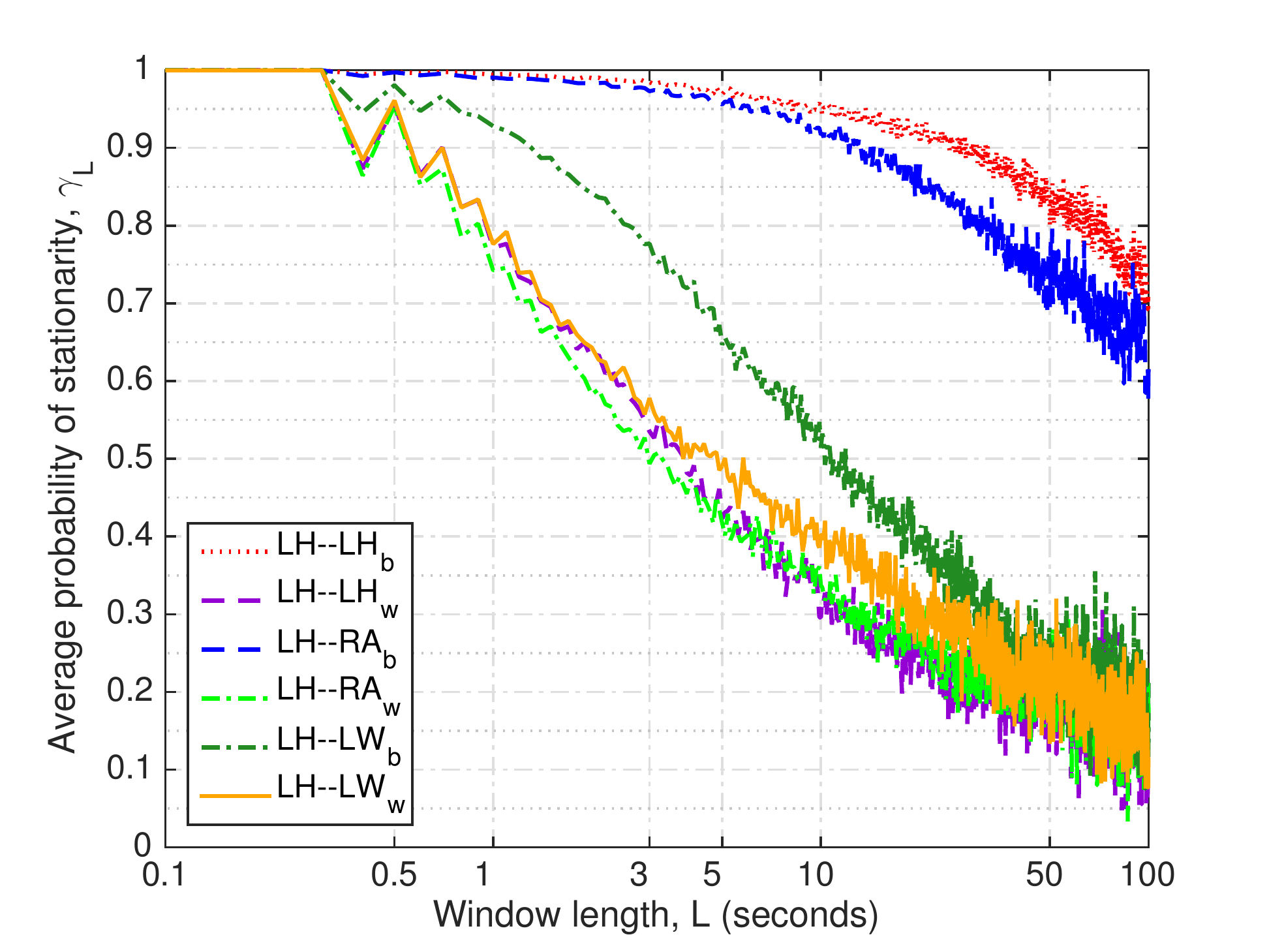}}
\caption{B--F hypothesis test for average probability of stationarity with $c\ell = 0.95$ across different body-to-body links, i.e., L. hip to L. hip (LH-LH), L. Hip to R. upper Arm (LH-RA), L. Hip to L. Wrist (LH-LW). Subscript `b' and `w' imply the best and worst case, respectively.}
\label{BF_b_w}
\end{figure}

The average probability of stationarity for the B--F hypothesis test averaged over $10$ BANs for different body-to-body links and for best/worst case links are shown in Figs. \ref{BF_all} and \ref{BF_b_w}, respectively. It can be seen from Fig. \ref{BF_all} that, the best stationarity condition persists for hub-to-hub links and the curves for hub-to-sensor links (i.e., LH--RA, LH--LW) are shifted to the right of those of the ANOVA test (Fig. \ref{anova_all}), indicating better stationarity characteristics. For example, there is $80\%$ probability for the LH--LH links to satisfy the WSS assumption for a window length greater than $10$ seconds with $99\%$ confidence level and up to $8$ seconds with $95\%$ confidence level. Also, the LH--RA and LH--LW links show $80\%$ probability of being stationary for greater than $8$ s and greater than $5$ seconds window lengths, respectively, with $99\%$ confidence. In Fig. \ref{BF_b_w}, the best-case LH--LH and LH--RA links show similar results to the ANOVA tests. Although, the worst-case curves show slightly improved results, the probability of stationarity slowly decreases to less than $50\%$ for window lengths greater than $3$ seconds.

\subsection{Distribution Consistency}
\begin{figure}[!t]
\centering{\includegraphics[width=\figwidth]{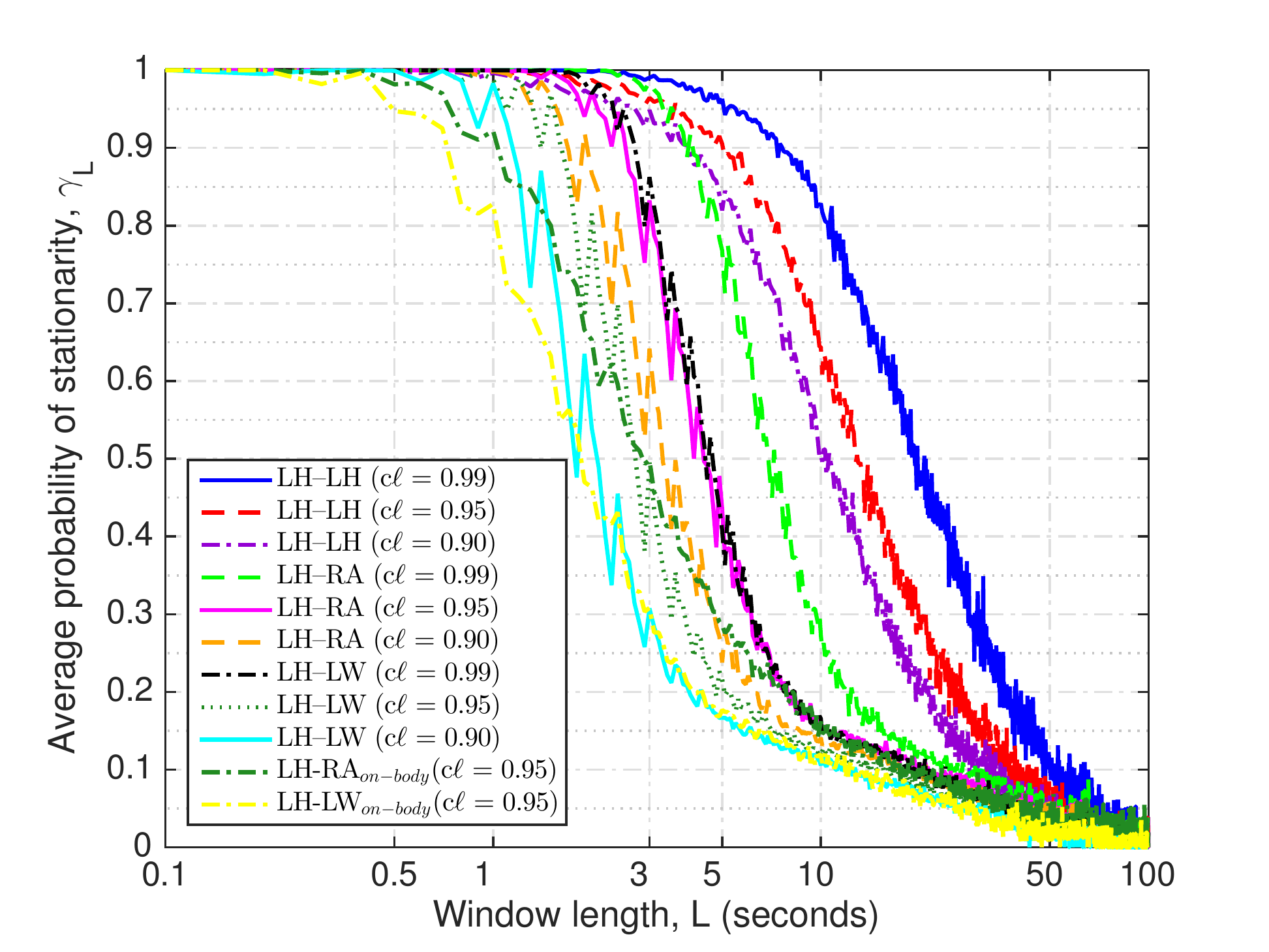}}
\caption{K-S hypothesis test for average probability of stationarity across different body-to-body links, i.e., L. hip to L. hip (LH-LH), L. Hip to R. upper Arm (LH-RA), L. Hip to L. Wrist (LH-LW) and on-body links, i.e., LH-RA$_{on-body}$, LH-LW$_{on-body}$ over $10$ subjects.}
\label{KS}
\end{figure}

\begin{figure}[!t]
\centering{\includegraphics[width=\figwidth]{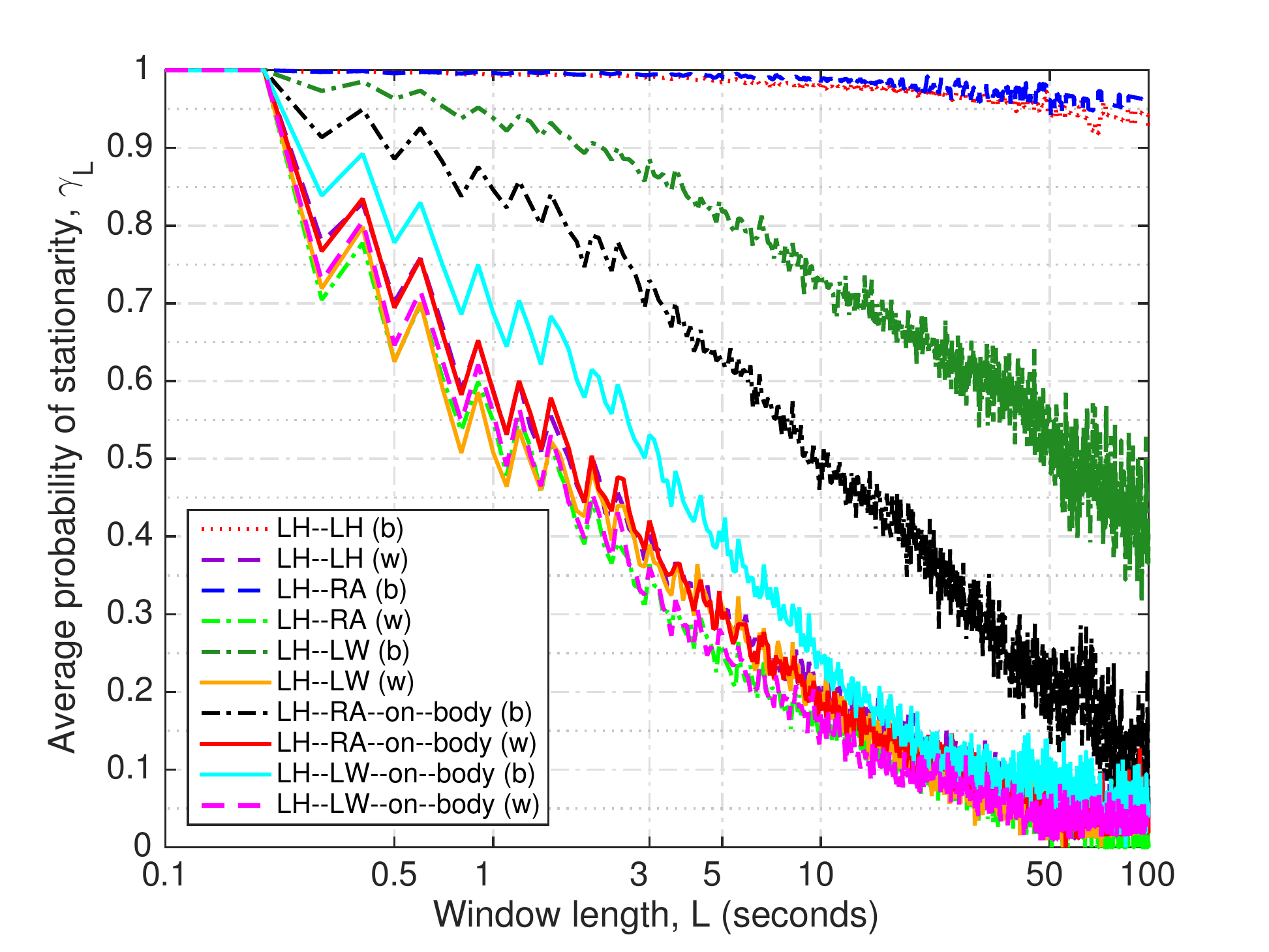}}
\caption{K-S hypothesis test for average probability of stationarity with $c\ell = 0.95$ across different body-to-body links, i.e., L. hip to L. hip (LH-LH), L. Hip to R. upper Arm (LH-RA), L. Hip to L. Wrist (LH-LW) and on-body links, i.e., LH-RA-on-body, LH-LW-on-body. `b' and `w' imply the best and worst case, respectively.}
\label{KS_b_w}
\end{figure}

The average probability of stationarity for different window lengths from K-S test over $10$ BANs with different body-to-body links is shown in Fig. \ref{KS}. The same process is applied for two types of on-body links, i.e., Left-Hip to Right-upper-Arm and Left-Hip to Left-Wrist over $10$ BANs to compare with B$2$B links. From Fig. \ref{KS}, it can be seen that, for window lengths $\geq$ $3$ s and $c\ell = 0.95$, there is less than $50\%$ chance of satisfying the null hypothesis over the total period for the on-body links, which can be considered as non-stationary behavior. On the other hand, the hub-to-hub (Left-Hip to Left-Hip) links between different BANs provides the best outcome with $90\%$ and $60\%$ probability of stationarity for window lengths of $5$ s and $10$ s, respectively, with $c\ell = 0.95$. The probability of stationarity decreases when considering the B$2$B links between the hubs and sensors (e.g., LH--RA, LH--LW) of different BANs, yet there is $70\%$ chance of being stationary for LH--LW and LH--RA with windows of more than $3$ s and $5$ s duration, respectively, with $c\ell = 0.99$.
The average probability of stationarity for the K-S test with single body-to-body links and on-body links for the best and worst conditions with $95\%$ confidence level is shown in Fig. \ref{KS_b_w}. As can be seen from this figure, the best-case hub-to-hub (LH--LH) and hub-to-sensor (LH--RA) links show excellent probability of stationarity for a large window length of $100$ s, whereas the probability of stationarity slowly decreases for the LH--LW link. However, in worst-case scenarios the B$2$B links depict non-stationary behavior for window lengths longer than $3$ s, whilst in best-case scenario, the on-body links show lower probability of satisfying the null hypothesis than B$2$B links.

The results from different hypothesis tests with hub-to-hub links is summarized in Table \ref{table_result}.
\begin{table}[!t]
\centering
\caption{The probability of stationarity for hub-to-hub links (LH--LH) over $10$ BANs, with $95\%$ confidence level}
\vspace{-0.5em}
\label{table_result}
\begin{tabular}{|c|c|c|c|c|}\Xhline{1pt}
\multirow{ 2}{*} {\begin{tabular}{@{}c@{}}Window\\length\end{tabular}} & \multirow{ 2}{*} {Hypothesis Test} & \multicolumn{ 3}{c|}{Probability of Stationarity}\\\cline{3-5}
 & & Average & Best-case & Worst-case\\\Xhline{1pt}
\multirow{ 3}{*} {$5$ s} & ANOVA Test & $0.7$ & $0.97$ & $0.3$\\\cline{2-5}
  & B--F Test & $0.9$ & $0.97$ & $0.4$\\\cline{2-5}
  & K--S Test & $0.9$ & $0.99$ & $0.3$\\\Xhline{1pt}
\multirow{ 3}{*} {$10$ s} & ANOVA Test & $0.3$ & $0.95$ & $0.3$\\\cline{2-5}
  & B--F Test & $0.6$ & $0.95$ & $0.3$\\\cline{2-5}
  & K--S Test & $0.6$ & $0.98$ & $0.2$\\\Xhline{1pt}
\end{tabular}
\end{table}

\vspace{0.5em}\textbf{Proposition 1} The WSS assumption for the body-to-body channels from Left-Hip to Left-Hip (LH--LH) can be held for window lengths of at least $5$~s.

\vspace{0.5em}\emph{Proof:} From the outcome of different hypothesis tests indicated in Table \ref{table_result}, the WSS assumption can not be rejected for the body-to-body channels from Left-Hip to Left-Hip (LH--LH) for window lengths of at least $5$ seconds (sampling frequency $20$ Hz) as there is up to $90\%$ chance (on average) that these links satisfy the null hypothesis within that duration. Also, in the best-case scenario, this duration can increase, e.g., $10$ s (with $95\%$ probability of being stationary).

\vspace{0.5em}\textbf{Remark 1} From the hypothesis test results, it appears that, the WSS assumption for body-to-body channels can depend on the on-body sensor locations. For example, the hub-to-hub (Left-Hip to Left-Hip) links show better probability of being stationary than hub-to-sensor (e.g., Left-Hip to Right-upper-Arm, Left-Hip to Left-Wrist) links.

\vspace{-0.04em}\subsection{Power Spectral Variation}
We investigate the variation in short-time power spectral coefficients \cite{basu2009nonparametric} of the B$2$B channels in windowed data segments over time, where we estimate the variance of multi-taper power spectral density (PSD) of specific data segments (e.g., $5$s, $10$s) over the whole channel.
\begin{equation}
{\hat{S_k}}^\xi (f) = \Bigg\lvert \sum_{t=1}^{L} g_k(t) X_\xi(t)e^{-i2\pi ft/L} \Bigg\rvert ^2 \qquad \xi = [1,2,...,M]
\end{equation}
where ${\hat{S_k}}^\xi (f)$ is the $k^{th}$ eigenspectrum found from the absolute square of the Short Time Fourier Transform (STFT) of window length $L$ and $g_k(t)$ is the $k^{th}$ rectangular window/taper from the discrete prolate spheroidal (Slepian) sequences of length $L$. $M$ is the number of windows over the whole channel ($M = N_c/L$ where $N_c$ is the length of the whole channel).
\begin{equation}
{\hat{S_\xi}} (f) = \frac{1}{K}\sum_{k=0}^{K-1} {\hat{S_k}}^\xi (f)
\end{equation}
where $K$ is the total number of the discrete prolate spheroidal sequences and ${\hat{S_\xi}} (f)$ is the multi-taper PSD estimation, which is the average of the $K$ modified periodograms.
\begin{equation}
{\hat{S}} (f) = \frac{1}{M}\sum_{\xi=1}^{M} \hat{S_\xi} (f)
\end{equation}
where ${\hat{S}} (f)$ is the average of PSD for $M$ windows over the whole channel and the variance of PSD with window length $L$ over the whole channel is
\begin{equation}
\vartheta_L = \frac{1}{M}\sum_{\xi=1}^{M} \bigg({\hat{S_\xi}} (f) - {\hat{S}} (f)\bigg)^2
\end{equation}
The power spectral variation ($V_L$) with window length $L$ over the whole channel is measured as follows,
\begin{equation}
V_L = \frac{1}{LM}\sum_{t=1}^{L} \sum_{\xi=1}^{M} \bigg({\hat{S_\xi}} (f) - {\hat{S}} (f)\bigg)^2
\end{equation}
The amount of $V_L$ would be $0$ when the channel is stationary over the window length $L$ \cite{basu2009nonparametric}. We found negligible amount of $V_L$ for $L =$ $5$ s, $10$ s with several B$2$B links, which implies that the channels can possess WSS characteristics.

\vspace{0.5em}\textbf{Remark 2} The short-time spectral coefficients of body-to-body channels show negligible variation over time by applying time-frequency analysis over windowed data segments. The channels hold the WSS assumption for segments with at least $5$ s duration with sampling rate of $20$ Hz (and can be larger, e.g., $10$ s with the same sampling rate).

\section{Conclusion}
In this paper, we investigated the wide-sense-stationarity (WSS) of body-to-body (B$2$B) links with the null hypothesis significance test (NHST) for different test statistics (difference between mean, variance and empirical distribution) with parametric and non-parametric approaches (i.e., ANOVA, Brown--Forsythe, Kolmogorov--Smirnov test). We have also examined the difference in auto-covariance in the frequency domain by estimating the variation in power spectrums. We have shown that the hub-to-hub links between different BANs can satisfy the WSS assumption with $95\%$ confidence level for a window length of at least $5$ seconds in up to $90\%$ of cases over the total period. This WSS region can be longer than that (e.g., up to $50$ s) in a best-case scenario for hub-to-hub links. It was also shown that the probability of satisfying the WSS assumption can depend on sensor locations as hub-to-hub links show better probability of satisfying the null hypothesis than hub-to-sensor links, which will be further investigated with data collected from different environments and sampling rates in future work. Overall the experimental results here indicate that the WSS assumption is held for body-to-body channels for segments of at least $5$ seconds duration (with $100$ samples). This is useful for modeling, forecasting and system design (e.g., channel state prediction, predictive decision making) in practical body-to-body networks, in contrast to BANs (with on-body channels) where such time for WSS does not hold.
%\section*{Acknowledgment}


%\bibliographystyle{IEEEtran}
\vspace{-0.5em}
\bibliographystyle{IEEEtran}
\bibliography{Reference}

\end{document}